# Submicrometer focusing of isolated attosecond XUV pulses approaching $10^{16}$ W/cm²

**Kotaro Imasaka,**[1,†] **Dianhong Dong,**[1,2,3,†] **Natsuki Kanda,**[1,†] **Bing Xue,**[1,2] **Satoru Egawa,**[4,5] **Takuya Hosobata,**[4] **Yutaka Yamagata,**[4] **Yasuo Nabekawa,**[1] **and Eiji J. Takahashi**[1,*]

[1]*Extreme Photonics Research Team, RIKEN Center for Advanced Photonics, RIKEN, 2-1, Hirosawa, Wako, Saitama, 351-0198, Japan*
[2]*State Key Laboratory of Ultrafast Optical Science and Technology, Xi'an Institute of Optics and Precision Mechanics, Chinese Academy of Sciences, Xi'an, 710119, China*
[3]*University of Chinese Academy of Sciences, Beijing 100049, China*
[4]*Ultrahigh Precision Optics Technology Team, RIKEN Center for Advanced Photonics, RIKEN, 2-1, Hirosawa, Wako, Saitama, 351-0198, Japan*
[5]*Research Center for Advanced Science and Technology, The University of Tokyo, 4-6-1, Komaba, Meguro, Tokyo, 153-8904, Japan*
[†]*These authors contributed equally.*
*\*ejtak@riken.jp*

We demonstrate submicrometer focusing of isolated attosecond pulses (IAPs) in the extreme ultraviolet (XUV) region using a custom ellipsoidal mirror. The obtained focal spot sizes were verified using knife-edge measurements with a sharp silicon edge, confirming reproducible dimensions down to 0.46 μm × 0.36 μm (FWHM), approaching the diffraction limit. Focusing a 1.1-GW tabletop IAP source yields a peak intensity of $3 \times 10^{15}$ W/cm², and a realistic pathway toward $10^{16}$ W/cm² is obtained by optimizing the beamline throughput. These results establish a practical route toward attosecond nonlinear optics in both gas and solid phases, driven by intense XUV fields.

## 1. Introduction

Attosecond pulses generated via high-order harmonic generation (HHG), including both attosecond pulse trains (APTs) and isolated attosecond pulses (IAPs), have become novel tools in ultrafast science [1,2], enabling direct access to electron dynamics on their natural timescale [3–7]. Over the past two decades, following the first experimental demonstration of IAP generation [8], the development of reliable IAP gating techniques, including polarization gating and double optical gating schemes [9,10], has enabled numerous studies on time-resolved electron dynamics in gases, largely within the linear-response regime [11–13]. More recently, advances in waveform-synthesized driving fields have drastically increased the pulse energy of IAPs, pushing their peak power into the gigawatt (GW) regime and opening the door to attosecond nonlinear optics [14,15].

Despite progress in IAP generation and control, achieving sufficiently high peak intensity for nonlinear applications remains experimentally challenging. To the best of our knowledge, experimentally demonstrated cases remain limited and have thus far been restricted to gaseous systems [15]. In current HHG-based attosecond sources, the available peak power and the difficulty in achieving sufficiently tight focusing limit the accessible interaction intensity with matter in experiments.

HHG-based attosecond sources exhibit excellent spatial coherence, enabling tight focusing into extremely small focal volumes. Thus, tight focusing provides a direct route to increase the interaction intensity without requiring further shortening of the attosecond pulse duration. This approach has been successfully demonstrated in other short-wavelength sources. In large-scale facilities, such as X-ray free-electron lasers (XFELs), which typically deliver femtosecond-duration pulses, submicrometer focusing has been demonstrated, and peak intensities exceeding $10^{16}$–$10^{20}$ W/cm² have been achieved [16–18]. Furthermore, attosecond pulses from XFELs have recently been reported [19,20], and experimental demonstrations combining attosecond-duration pulses and tightly focused beams have recently emerged [21]. However, XFEL sources are typically only available at large-scale user facilities with limited accessibility. These considerations motivate the development of compact tabletop sources capable of generating intense attosecond fields. In this context, HHG-based sources are highly promising candidates for realizing such a platform, providing an accessible route to high-intensity attosecond science.

Within the HHG-based framework, submicrometer focusing has been demonstrated for APTs [22,23]. Furthermore, peak intensities of the order of $10^{14}$ W/cm² have been reported for HHG beams focused to several micrometers in the extreme ultraviolet (XUV) region [24,25]. However, APTs comprise a sequence of attosecond bursts separated by half an optical cycle, and this intrinsic temporal structure complicates ultrafast measurements, particularly in pump–probe schemes that require a well-defined temporal excitation. Consequently, IAPs are generally preferred for experiments aiming to probe electron dynamics on the attosecond timescale. Therefore, achieving submicrometer focusing for IAPs is an important step toward achieving high peak intensities while preserving a well-defined temporal waveform of a single attosecond burst. However, the optical and metrological techniques required to achieve such focusing and accurately determine the focal size for IAPs have not been demonstrated.

In this study, we demonstrate submicrometer focusing of IAPs in the XUV region. By designing and fabricating a custom ellipsoidal focusing mirror with sufficiently small surface roughness, we identify the key conditions required for submicrometer focusing of HHG-based IAPs. APTs are initially used as a stable benchmark to evaluate the focusing performance of the optical system. Robust knife-edge measurements, which are widely used for focused HHG beam characterization in the XUV regime [26], were performed using a sharp silicon edge. These measurements demonstrate reproducible submicrometer focal spots before the measurement is extended to IAPs within the same optical configuration. For a 1.1-GW tabletop IAP source, the measured focal size corresponds to a peak intensity of $3.0 \times 10^{15}$ W/cm² and a peak fluence of 0.71 J/cm². These values indicate a realistic pathway toward the $10^{16}$-W/cm² regime through improved optical throughput in the present beamline. To the best of our knowledge, this study is the first experimental demonstration of submicrometer focusing of IAPs with quantitatively validated beam-size evaluations and achieves the highest focusing intensity reported for IAPs, marking a significant milestone in high-intensity attosecond science.

These results establish a focusing platform capable of generating isolated attosecond XUV fields with both attosecond temporal resolution and experimentally verified high peak intensities at the focus. Such tightly confined attosecond fields provide a promising route toward nonlinear XUV interactions. Furthermore, the demonstrated focusing capability is expected to provide a basis for future attosecond pump–probe experiments using well-characterized isolated attosecond fields.

## 2. Methods

### 2.1 Experimental setup for XUV focusing

For the systematic characterization of the focusing optics, APTs were generated by driving HHG with an 800-nm pump pulse alone. APTs provide a higher photon flux and more stable operation than waveform-synthesized IAP configurations. This enables the rapid and reproducible evaluation of the intrinsic focusing performance of an optical system, providing a reliable reference for subsequent IAP measurements despite differences in photon-energy distribution. The system was then operated using a three-color synthesized driving field and a Zr thin-film filter to spectrally select the cutoff region of the HHG spectrum (corresponding to photon energies above approximately 60 eV). This setup generated IAPs, allowing the investigation of the achievable peak intensity. The relevant spectral conditions for these measurements are summarized in Supplementary Note S1 (Fig. S1).

The experimental setup for HHG and XUV focusing is illustrated in Fig. 1(a), which shows the overall HHG beamline, including optical waveform synthesis, HHG in a gas cell, and XUV beam transport. The IAPs used in this study were generated via HHG driven by a three-channel optical waveform synthesizer, as previously reported [14,27]. Additional details of the IAP generation scheme and laser parameters are provided in Supplementary Note S1. After HHG in the gas cell, the residual driving laser light is significantly attenuated by NbN-coated mirrors [28] and reflected from the surface of a 500-nm-thick Al thin-film filter. The reflected infrared light is imaged onto a position-sensitive detector (PSD) located at a plane conjugate to the XUV focus. This setup provides feedback to a PZT-actuated steering mirror to stabilize the XUV focal position, as detailed in Sec. 2.2. The generated XUV beam is transported through the beamline using NbN-coated mirrors and thin-film filters, including Al and, for IAP operation, Zr. The beam passes through a shutter before entering the focusing stage. The ellipsoidal focusing mirror was designed to form an image of the HHG source, which is located approximately 4 m upstream, onto the focal plane with a focal length of 60 mm, resulting in an incidence angle of approximately 45°. Supplementary Note S2 provides additional details of the mirror design and associated wave-optical simulations, which form the basis for these surface figure error requirements. The mirror substrate was fabricated and evaluated by Natsume Optical Corp. The measured surface quality confirms that the surface figure error within the illuminated region satisfies these requirements. The detailed surface characterization results are summarized in Supplementary Note S3. Subsequently, a Mo/Si multilayer coating was deposited on the mirror surface by NTT Advanced Technology Corp. The multilayer was designed to tailor the spectral response to the cutoff region of the harmonic spectrum, providing high reflectivity over the cutoff region while suppressing lower-order harmonics (Fig. S1).

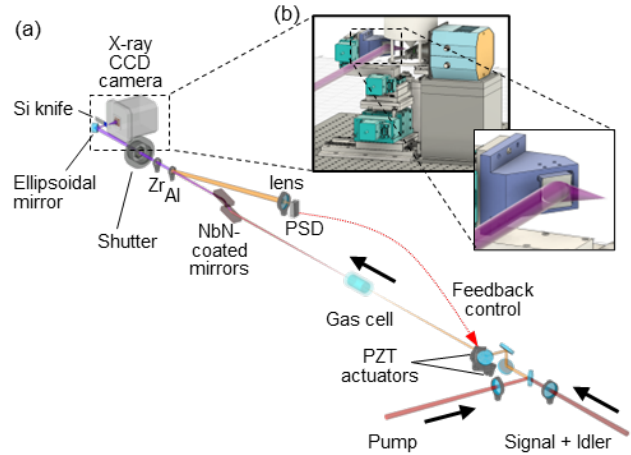

Fig. 1. Experimental setup for HHG and XUV focusing. (a) Schematic of HHG beamline. The system comprises waveform synthesis, HHG in a gas

cell, and XUV beam transport. (b) Enlarged view of the focusing region. The incident beam is drawn thicker than the scale for clarity.

Fig. 1(b) shows the focusing configuration, including the focusing mirror and focal diagnostics. The ellipsoidal focusing mirror is mounted on a five-axis precision stage with three translational and two rotational degrees of freedom. The mirror angle is optimized using a Foucault test by monitoring the XUV beam profile transmitted past the knife edge with an X-ray charge-coupled device (CCD) camera. The X-ray CCD images for APT and IAP shown in Supplementary Figs. S1(d) and (e) exhibit comparable spatial beam profiles in terms of size and shape. Additional details of the beamline configuration are provided in Supplementary Note S4. The focal diagnostics, including knife-edge measurements and focal-position stabilization, are explained in the next section.

## 2.2 Knife-edge measurements and focal-position stabilization

Focal-spot characterization was performed using a knife-edge method with a sharp Si edge and an X-ray CCD camera. A silicon knife edge was mounted on a three-axis closed-loop piezo translation stage with nanometer-scale positioning capability. To suppress the transmission of XUV radiation through the thin edge region, the silicon knife edge was coated with a 130-nm-thick Ni layer (with a transmission below approximately 0.3% in the present photon-energy range [29,30]). The XUV intensity downstream of the knife edge was recorded as a function of its position using an X-ray CCD camera. The spatially integrated signal was used as the measure of intensity. A mechanical shutter was used to block the XUV beam during the CCD readout. An additional Al thin-film filter mounted on the CCD entrance ensured that only XUV radiation reached the detector. Multiple exposures were acquired at each knife position to improve statistical reliability. The acquisition parameters were adjusted according to the photon flux and stability of the operating conditions, as summarized in Supplementary Note S5. The full width at half maximum (FWHM) of the focal spot was obtained by fitting the resulting knife-edge curve with an error function, corresponding to the integral of a Gaussian intensity profile.

The focal-position stabilization system shown in Fig. 1 was used to maintain the XUV focal position during acquisition durations of several tens of minutes. The reflected infrared beam was focused by a lens onto a PSD, providing a reference signal for the XUV focal position. Fluctuations in the driving beam translate into corresponding fluctuations in the XUV beam because high-order harmonics are generated collinearly with the driving laser field.

The PSD signal, synchronized to the 10-Hz repetition rate of the driving laser, was converted into a smooth signal using a resistor–capacitor circuit with a cutoff frequency on the order of 1 Hz and used as an error signal relative to a predefined reference position. Feedback control was applied to a piezo-actuated mirror upstream of the HHG interaction region, enabling the stabilization of the XUV focal position.

The performance of the focal-position stabilization system was evaluated by monitoring the PSD-derived focal-position signal in the horizontal (X) and vertical (Y) directions (Fig. 2). In the absence of active feedback, slow long-term drifts with a characteristic timescale of approximately 1 h were observed, resulting in standard deviations of 2.76 μm (X) and 3.52 μm (Y). Engaging the feedback loop significantly suppressed the fluctuations, reducing the standard deviations to 0.79 μm (X) and 0.96 μm (Y).

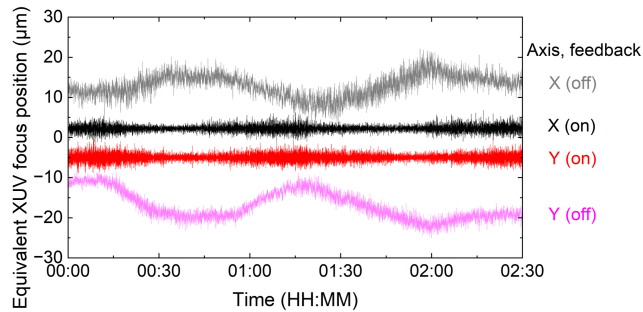

Fig. 2. Performance of focal-position stabilization system. Temporal traces of the PSD-derived focal-position signal in the horizontal (X) and vertical (Y) directions with and without feedback.

Long-term drift suppression is essential for reliable knife-edge measurements because scans typically require tens of minutes; otherwise, focal-position drift during the scan would lead to a systematic overestimation or underestimation of the focal spot size. Even in the presence of feedback, residual focal-position fluctuations remain, with a standard deviation of approximately 1 μm. Such fluctuations can still introduce a systematic overestimation of the focal spot size when conventional knife-edge data averaging is used. To mitigate this effect, the median of repeated measurements at each knife-edge position was used as the robust estimator of the focal spot size. The quantitative validity of this procedure and its impact on the extracted beam size are discussed in detail in Section 3.2.1.

## 3. Results and Discussion

### 3.1 Performance of XUV focusing platform characterized with APTs

The entire XUV focusing platform was initially characterized using APTs to assess its intrinsic focusing performance. In this configuration, HHG was driven solely by the 800-nm pump pulse without using a Zr thin-film filter for spectral selection, whereas the beamline alignment and focusing geometry remained identical to those used for the IAP operation.

Figs. 3(a) and (b) show representative knife-edge measurements obtained using APTs in the horizontal (X) and vertical (Y) directions, respectively. The small offset observed in the knife-edge signals originates from background contributions but does not affect the extracted beam size. The shot-to-shot fluctuations visible in the raw accumulated signals (gray cross markers) are effectively suppressed by the median-based estimator, resulting in smooth transitions. The median-processed XUV intensity at each knife position (black circles) is well-fitted by error functions (red lines), with their derivatives corresponding to Gaussian beam profiles (the insets). This agreement confirms the robustness of the beam-size extraction procedure. The details of the acquisition parameters are provided in Supplementary Note S5.

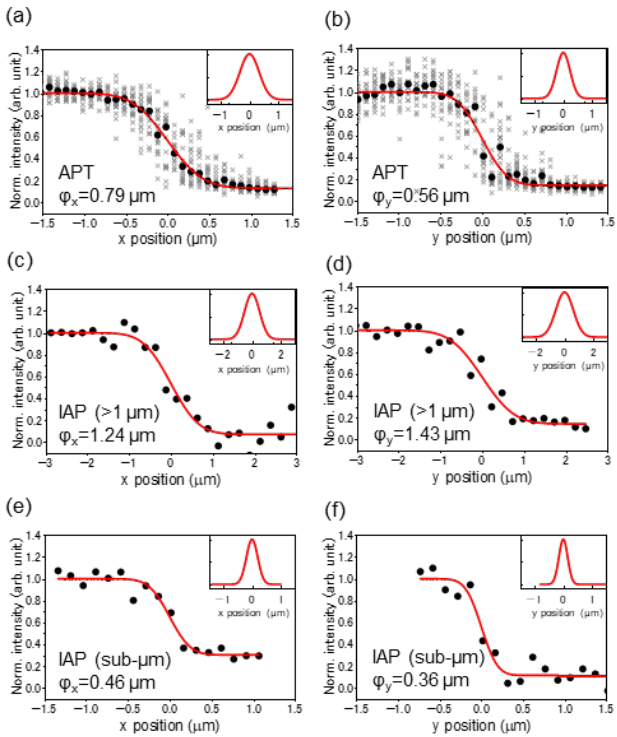

Fig. 3. Knife-edge characterization of focused XUV beam. (a) and (b) APT measurements in the horizontal (X) and vertical (Y) directions, respectively. (c) and (d) IAP measurements showing spot sizes slightly above 1 μm. (e) and (f) IAP measurements demonstrating submicrometer focusing. The gray cross markers represent raw accumulated signals, the black circles represent the median-processed intensity values, and the red lines represent the error function fits. The insets show the corresponding reconstructed beam profiles obtained from the derivative of the fitted error functions, representing Gaussian beam profiles.

The focal spot sizes retrieved from the error function fits are 0.79 ± 0.06 μm (X) and 0.56 ± 0.10 μm (Y) in terms of FWHM. Although the photon-energy distribution of the APT spectrum differs from that of the IAP spectrum, the shorter wavelength components in the IAP spectrum are expected to enable tighter diffraction-limited focusing assuming the same optical configuration and comparable incident beam sizes. These results demonstrate that the optical system can achieve tightly confined XUV beams and establish a reliable reference for subsequent IAP measurements.

### 3.2 Submicrometer focusing of IAPs

IAP generation via cutoff selection employs a Zr thin-film filter and the reflectivity of the Mo/Si multilayer focusing mirror. The acquisition parameters for the knife-edge measurements were adjusted to ensure stable signal acquisition during the IAP operation (Supplementary Note S5).

Representative knife-edge results obtained using IAPs are shown in Figs. 3(c)–(f). Using the alignment established with APTs as a starting point, we finely adjusted the mirror position and angle. Two distinct focusing regimes are observed depending on the illuminated region of the mirror surface: one with spot sizes slightly above 1 μm (Figs. 3(c) and (d)) and another achieving submicrometer confinement (Figs. 3(e) and (f)).

In the latter case, the smallest focal spot sizes reach 0.46 ± 0.13 μm in the horizontal direction and 0.36 ± 0.12 μm in the vertical direction (FWHM), indicating submicrometer focusing. For the IAP measurements (Figs. 3(c)–(f)), three independent scans in both directions yielded consistent results, confirming the reproducibility of the focusing behavior (see Supplementary Note S5). In contrast, the larger spot sizes observed in Figs. 3(c) and (d) indicate a deviation from the expected submicrometer focusing performance. The origin of this behavior is further discussed in a later section.

### 3.2.1 Robust evaluation of beam size under residual fluctuations

In this section, we examine the performance of our statistical analysis based on the median values from the knife-edge measurements. Even with the focal-position stabilization described in Section 2.2, the residual motion of the XUV focus on the order of 1 μm remains, as inferred from the PSD-referenced signal. Beam-position fluctuations can result in deviations in the detected intensity, especially when the knife edge is positioned near the beam edge, and can introduce systematic bias in the retrieved beam size when conventional averaging is used.

Fig. 4(a) illustrates numerical simulations of this effect for a representative knife position of −0.5 μm, where random beam-position shifts with a standard deviation of 1 μm are modeled using 50 realizations. The resulting distribution of transmitted intensities becomes significantly asymmetric, causing the mean intensity to deviate significantly from the ideal value defined in the absence of beam-position fluctuations. Consequently, beam reconstruction using the mean results in a substantial overestimation of the beam size, as evident from the fitted profiles in Fig. 4(b) (see also Supplementary Note S6).

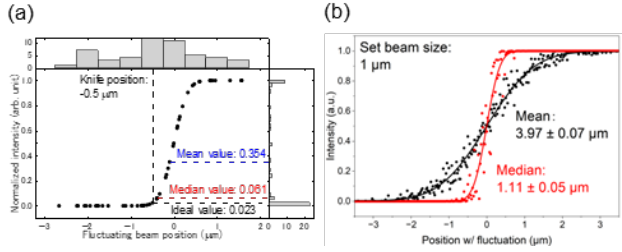

Fig. 4. Influence of statistical estimators on knife-edge beam reconstruction under residual focal fluctuations. (a) Simulated transmitted intensities as a function of beam position at a representative knife position, along with the corresponding histograms of beam-position fluctuations (top) and transmitted intensities (right). (b) Reconstructed beam profiles using mean and median estimators. Markers indicate the reconstructed data points, and solid lines represent error function fits.

To overcome this bias, the median is used as a robust estimator. Because the underlying beam-position fluctuations occur without directional preference, the transmission corresponding to the most probable overlap between the beam and the knife edge is better represented by the median. In contrast to the mean, the median remains close to the ideal transmission value even near the beam edge (Fig. 4(a)), resulting in a reconstructed beam size that remains close to the intrinsic beam size (Fig. 4(b)). These results demonstrate that under realistic experimental fluctuations, a median estimator provides a reliable and physically justified measure of the focal spot size. Short-term focal-position drifts during knife-edge scans were also mitigated using a dedicated scanning protocol (Supplementary Note S7).

### 3.2.2 Effect of local mirror surface height variations on focal spot size

We investigate the origin of the variation in focal spot size between the slightly-above-1-μm cases in Figs. 3(c) and (d) and the submicrometer cases in Figs. 3(e) and (f). The illuminated region on the mirror was varied by shifting the incident beam (1–2 mm in diameter on the mirror surface) across distances comparable to its size. Because this beam size is significantly smaller than the full-aperture region shown in Figs. S3(a) and (b), the observed variation in focal spot size indicates that local surface variations within the illuminated region primarily influence focusing performance. Surface variations at submillimeter spatial wavelengths are expected to play a dominant role, because their spatial scale is smaller than the beam size on the mirror and can introduce phase distortions in the reflected wavefront.

To evaluate sensitivity to local surface variations, wave-optical simulations were performed using the surface height map shown in Fig. S3(c), which captures these submillimeter components. Illumination position variations were approximated by scaling the magnitude of local surface height variations based on the assumption that different mirror surface regions share similar spatial structures but exhibit different levels of surface height variations at submillimeter spatial wavelengths (Supplementary Note S8).

Fig. 5 shows the resulting focal intensity distributions. The projection profiles within the two-dimensional (2D) maps were obtained by integrating the intensity distribution along the orthogonal direction. The corresponding knife-edge responses are used to reconstruct the beam sizes. Although the 2D maps exhibit a well-defined central peak, the surrounding intensity distribution extends beyond the main lobe, particularly for increased surface variations. Consequently, the corresponding knife-edge signals include contributions from the extended distribution, significantly increasing the reconstructed beam size. For the surface map corresponding to the measured roughness level (1×), the reconstructed knife-edge profiles yield focal spot sizes of 0.57 μm (X) and 0.59 μm (Y), consistent with submicrometer focusing. When the surface variation is increased to twice the measured value, the reconstructed beam sizes increase to 1.09 μm (X) and 3.29 μm (Y). For the case of three times the measured roughness, the focal profile becomes significantly distorted, and the reconstructed beam sizes further increase to 3.03 μm (X) and 10.29 μm (Y).

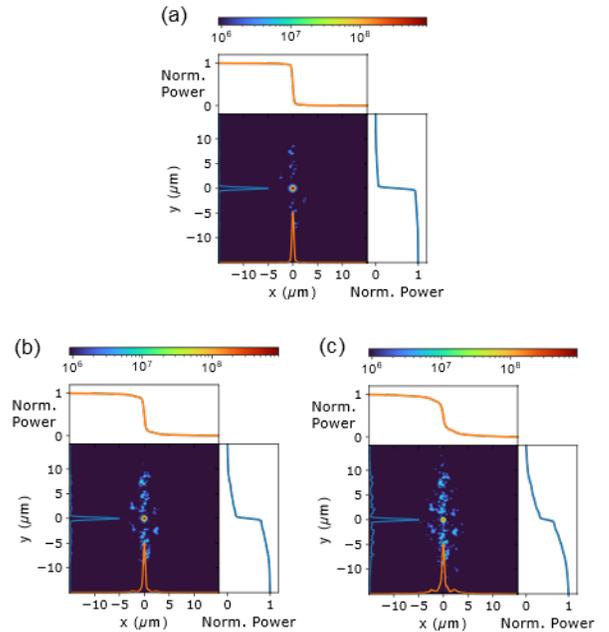

Fig. 5. Wave-optical simulations of the influence of local mirror surface variations on the focal spot size. (a) Measured surface map (1× roughness). (b) Surface map with twofold surface height variations. (c) Surface map with threefold surface height variations. For each case, the central panel shows the calculated 2D focal intensity distribution (logarithmic scale). Projection profiles (integrated along the orthogonal direction) are shown within the 2D maps (bottom and left axes). The upper and right panels show the corresponding knife-edge profiles used for beam-size reconstruction.

These results indicate that moderate increases in local surface variations at submillimeter spatial wavelengths enhance the effect of wavefront distortion, thereby altering the knife-edge transmission profiles and leading to an overestimation of the reconstructed focal spot size, readily degrading it to the micrometer scale. This behavior provides a natural interpretation of the contrast between the slightly-above-1-μm regime observed in Figs. 3(c) and (d) and the submicrometer regime shown in Figs. 3(e) and (f). These results suggest that unmeasured mirror surface regions (three measured areas of 1 mm × 1 mm within a 15 mm × 15 mm effective aperture) may exhibit greater surface height variations at submillimeter spatial wavelengths. The transition between the two focusing regimes primarily originates from spatial variations of the local mirror surface rather than alignment-related effects.

Combined with the statistical analysis in Section 3.2.1, these results support the conclusion that the observed submicrometer focusing reflects the intrinsic performance of the optical system.

### 3.3 Peak focusing intensity of IAPs

The peak intensity achievable in the present configuration is estimated from the measured focal spot sizes, along with the pulse energy and duration reported for the same IAP source [14]. The spectral throughput of the beamline is evaluated based on literature data for optical elements [28–30]. Under the assumption of Gaussian spatial and temporal profiles, the peak intensity is given by $I_0 = 8\sqrt{(\ln 2/\pi)^3} \times \varepsilon/(\varphi_x \varphi_y \Delta t)$, where $\varepsilon$ denotes the pulse energy at the focus, $\varphi_x$ and $\varphi_y$ denote the focal FWHM beam sizes in the horizontal and vertical directions, respectively, and $\Delta t$ denotes the pulse duration (FWHM). Using the measured focal FWHM dimensions of 0.46 μm (X) and 0.36 μm (Y), the resulting peak intensity is estimated to be $3.0 \times 10^{15}$ W/cm².

A realistic increase in peak intensity can be expected by reducing optical losses. For example, operation with a single NbN reflection, removal of the 500-nm-thick Al filter used for stabilization, and the use of a thinner Zr filter are technically feasible modifications. In particular, replacing the 100-nm-thick Zr film with an approximately 40-nm-thick film while maintaining sufficient suppression of the low-photon-energy components will increase the transmitted pulse energy. Under these conditions, the peak intensity is expected to reach $3.1 \times 10^{16}$ W/cm².

In the context of HHG-based XUV focusing, previously reported peak intensities are typically on the order of $10^{14}$ W/cm² in APT-based focusing studies [24,25]. In contrast, in this study, the peak intensity reaches $3.0 \times 10^{15}$ W/cm² and is projected to extend to the $10^{16}$-W/cm² regime under reduced optical losses. This corresponds to an extension of the accessible intensity range by more than an order of magnitude compared with previous HHG-based focusing experiments. This introduces HHG-based attosecond sources into an intensity regime where nonperturbative nonlinear responses in solids have been observed using femtosecond XUV sources [18], while such responses remain largely unexplored for attosecond pulses. This provides access to the nonperturbative nonlinear regime in solids with IAPs in a tabletop platform. This corresponds to a peak fluence of approximately 0.71 J/cm², which exceeds typical fluence levels used in femtosecond laser processing (on the order of several tens of mJ/cm²) and suggests that attosecond-scale material processing is feasible. As previously discussed, this peak fluence can be increased to several J/cm² by optimizing the optical throughput in the present beamline.

A similar estimate can also be made for higher-photon-energy IAP sources previously reported by our group. In Ref. [27], high-order harmonics generated in neon provided pulse energies in the cutoff region on the order of a few tens of nJ. Subsequent all-optical frequency-resolved optical gating measurements revealed a pulse duration of 128 as at a photon energy of 107 eV [31]. Assuming comparable focusing conditions, these parameters correspond to peak intensities of the order of $10^{14}$ W/cm². Although operation in this photon-energy range requires spectral optics modifications, this estimate indicates that high-peak-intensity attosecond fields can be extended to broader spectral ranges.

### 4. Conclusion

The submicrometer focusing of IAPs was demonstrated using a custom-designed ellipsoidal multilayer mirror optimized for XUV operation. The smallest focal spots reached 0.46 ± 0.13 μm (horizontal) and 0.36 ± 0.12 μm (vertical) in FWHM, consistent with the near-diffraction-limited performance expected from wave-optical simulations of the mirror design. By combining robust statistical estimators for knife-edge analysis with controlled acquisition protocols and a detailed understanding of the mirror surface quality, we developed a beam-size evaluation approach that reliably reveals intrinsic focusing performance. For a GW-class tabletop IAP source, the experimentally demonstrated focal dimensions correspond to peak intensities of the order of $3.0 \times 10^{15}$ W/cm². In addition, straightforward reductions in optical losses indicate a realistic path toward the $10^{16}$-W/cm² regime. These results demonstrate that intense IAPs generated in tabletop HHG systems enable both attosecond temporal resolution and experimentally validated high peak intensities. This capability provides a practical platform for nonlinear attosecond optics in gases and solids, and it is expected to enable attosecond pump–probe experiments using tabletop HHG systems.

**Funding.** MEXT - Quantum Leap Flagship Program (JP-MXS0118068681); JSPS KAKENHI (JP23K13713, 26K17782).

**Acknowledgment.** We thank Masahiro Takeda of the Ultrahigh Precision Optics Technology Team for his assistance in fabricating and evaluating the prototype focusing mirror; Natsume Optical Corp. for fabricating the ellipsoidal focusing mirror substrate and providing the surface metrology data used in this study. This project was supported by the RIKEN TRIP initiative (Leading-edge semiconductor technology).

**Disclosures.** The authors declare no conflicts of interest.

**Data availability.** Data underlying the results presented in this paper are not publicly available at this time but may be obtained from the authors upon reasonable request.

**Supplemental document.** See Supplement 1 for supporting content.

**References**

1. P. B. Corkum, "Plasma perspective on strong field multiphoton ionization," Phys. Rev. Lett. **71**, 1994 (1993).
2. M. Lewenstein, Ph. Balcou, M. Yu. Ivanov, *et al.*, "Theory of high-harmonic generation by low-frequency laser fields," Phys. Rev. A **49**, 2117 (1994).
3. F. Calegari, G. Sansone, S. Stagira, *et al.*, "Advances in attosecond science," J. Phys. B **49**, 062001 (2016).
4. A. Baltuška, Th. Udem, M. Uiberacker, *et al.*, "Attosecond control of electronic processes by intense light fields," Nature **421**, 611 (2003).
5. M. Chini, K. Zhao, and Z. Chang, "The generation, characterization and applications of broadband isolated attosecond pulses," Nat. Photon. **8**, 178 (2014).
6. J. Itatani, J. Levesque, D. Zeidler, *et al.*, "Tomographic imaging of molecular orbitals," Nature **432**, 867 (2004).
7. L. Ortmann and A. Landsman, "Understanding attosecond streaking," Rep. Prog. Phys. **87**, 086401 (2024).
8. M. Hentschel, R. Kienberger, Ch. Spielmann, *et al.*, "Attosecond metrology," Nature **414**, 509 (2001).
9. I. J. Sola, E. Mével, L. Elouga, *et al.*, "Controlling attosecond electron dynamics by phase-stabilized polarization gating," Nat. Phys. **2**, 319 (2006).
10. H. Mashiko, S. Gilbertson, C. Li, *et al.*, "Double optical gating of high-order harmonic generation with carrier-envelope phase stabilized lasers," Phys. Rev. Lett. **100**, 103906 (2008).
11. M. Nisoli, P. Decleva, F. Calegari, *et al.*, "Attosecond electron dynamics in molecules," Chem. Rev. **117**, 10760 (2017).
12. N. Saito, H. Sannohe, N. Ishii, *et al.*, "Real-time observation of electronic, vibrational, and rotational dynamics in nitric oxide with attosecond soft x-ray pulses at 400 eV," Optica **6**, 1542 (2019).
13. Y. Ma, H. Ni, and J. Wu, "Attosecond ionization time delays in strong-field physics," Chin. Phys. B **33**, 013201 (2024).
14. B. Xue, K. Midorikawa, and E. J. Takahashi, "Gigawatt-class, tabletop, isolated-attosecond-pulse light source," Optica **9**, 360 (2022).
15. E. J. Takahashi, P. Lan, O. D. Mücke, *et al.*, "Attosecond nonlinear optics using gigawatt-scale isolated attosecond pulses," Nat. Commun. **4**, 2691 (2013).
16. H. Yumoto, H. Mimura, T. Koyama, *et al.*, "Focusing of X-ray free-electron laser pulses with reflective optics," Nat. Photon. **7**, 43 (2013).
17. H. Mimura, H. Yumoto, S. Matsuyama, *et al.*, "Generation of $10^{20}$ W cm$^{-2}$ hard X-ray laser pulses with two-stage reflective focusing system," Nat. Commun. **5**, 3539 (2014).
18. H. Motoyama, S. Owada, G. Yamaguchi, *et al.*, "Intense sub-micrometre focusing of soft X-ray free-electron laser beyond $10^{16}$ W cm$^{-2}$ with an ellipsoidal mirror," J. Synchrotron Rad. **26**, 1406 (2019).
19. J. Duris, S. Li, T. Driver, *et al.*, "Tunable isolated attosecond X-ray pulses with gigawatt peak power from a free-electron laser," Nat. Photon. **14**, 30 (2020).
20. P. Franz, S. Li, T. Driver, *et al.*, "Terawatt-scale attosecond X-ray pulses from a cascaded superradiant free-electron laser," Nat. Photon. **18**, 698 (2024).
21. I. Inoue, T. Sato, R. Robles, *et al.*, "Nanofocused attosecond hard x-ray free-electron laser with intensity exceeding $10^{19}$ W/cm$^2$," Optica **12**, 309 (2025).
22. K. Sakaue, H. Motoyama, R. Hayashi, *et al.*, "Surface processing of PMMA and metal nano-particle resist by sub-micrometer focusing of coherent extreme ultraviolet high-order harmonics pulses," Opt. Lett. **45**, 2926 (2020).
23. H. Motoyama, A. Iwasaki, H. Mimura, *et al.*, "Submicron structures created on Ni thin film by submicron focusing of femtosecond EUV light pulses," Appl. Phys. Express **16**, 016503 (2023).
24. B. Major, O. Ghafur, K. Kovács, *et al.*, "Compact intense extreme-ultraviolet source," Optica **8**, 960 (2021).
25. E. Sobolev, M. Volkov, E. Svirplys, *et al.*, "Terawatt-level three-stage pulse compression for all-attosecond pump-probe spectroscopy," Opt. Express **32**, 46251 (2024).
26. L. Le Déroff, P. Salières, and B. Carré, *et al.*, "Beam-quality measurement of a focused high-order harmonic beam," Opt. Lett. **23**, 1544 (1998).
27. B. Xue, Y. Tamura, Y. Fu, *et al.*, "A custom-tailored multi-TW optical electric field for gigawatt soft-X-ray isolated attosecond pulses," Ultrafast Sci. **2021**, 9828026 (2021).
28. Y. Nagata, Y. Nabekawa, and K. Midorikawa, "Development of high-throughput, high-damage-threshold beam separator for 13 nm high-order harmonics," Opt. Lett. **31**, 1316 (2006).
29. B. L. Henke, E. M. Gullikson, and J. C. Davis, "X-ray interactions: photoabsorption, scattering, transmission, and reflection at E= 50-30,000 eV, Z= 1-92," At. Data Nucl. Data Tables **54**, 181 (1993).
30. Henke optical constants database, https://henke.lbl.gov/optical_constants/filter2.html (accessed Sep. 11, 2025).
31. D. Dong, H. Wang, B. Xue, *et al.*, "Perturbed three-channel waveform synthesizer for efficient isolated attosecond pulse generation and characterization," Opt. Lett. **50**, 1461 (2025).

# SUBMICROMETER FOCUSING OF ISOLATED ATTOSECOND XUV PULSES APPROACHING $10^{16}$ W/CM$^2$:

## SUPPLEMENTAL DOCUMENT

### *Supplementary Note S1. Generation of isolated attosecond pulses*

The IAPs used in this study were generated via HHG in an argon gas medium driven by a three-channel optical waveform synthesizer, following a previously reported scheme [27]. The synthesized driving field comprised pump, signal, and idler pulses with central wavelengths of 800, 1350, and 2050 nm, respectively. The pulse durations of the pump, signal, and idler pulses were 30, 44, and 86 fs, with corresponding pulse energies of 21.3, 2.8, and 1.9 mJ, respectively. The system was operated at a repetition rate of 10 Hz. To ensure reproducible waveform synthesis, three stabilization schemes were incorporated: carrier-envelope phase stabilization of the 800-nm driving laser, active stabilization of the relative time delays among the three channels, and beam-pointing stabilization using the 800-nm beam to stabilize the HHG interaction position.

The high-harmonic radiation generated by the three-color driving field was spectrally filtered using a Zr thin film to transmit the cutoff region of the harmonic spectrum reaching ~70 eV while suppressing harmonics below ~50 eV, selecting IAPs. Under these conditions, the generated IAPs were characterized by a pulse duration of 226 as, a pulse energy of 0.24 μJ, and a peak power on the order of 1.1 GW, as previously reported [14].

The spectral characteristics of both APT and IAP used in this study were derived from spectrometer measurements combined with the spectral response of the beamline optics. Fig. S1(a) shows the resulting spectra corresponding to the XUV focus. The corresponding 2D spatial–spectral intensity distributions, obtained using an XUV spectrometer without the focusing optics, are shown in Figs. S1(b) and (c) for APT and IAP, respectively. These measurements provide the vertical divergence characteristics of the generated XUV beams in this configuration. The corresponding 2D beam profiles obtained with the focusing system in place using an X-ray CCD camera are shown in Figs. S1(d) and (e). The beam profiles observed at this stage are quantitatively similar for APT and IAP, with FWHM beam sizes of 2.1 mm × 1.5 mm and 1.9 mm × 1.3 mm, respectively. The difference between the two cases is approximately 10%–15%, indicating that comparable beam conditions are achieved for APT and IAP in the focusing experiment. A direct comparison of the beam profiles along the vertical direction obtained from the spectrometer and the X-ray CCD is shown in Figs. S1(f) and (g) for APT and IAP, respectively, demonstrating agreement of the divergence values within approximately 10%.

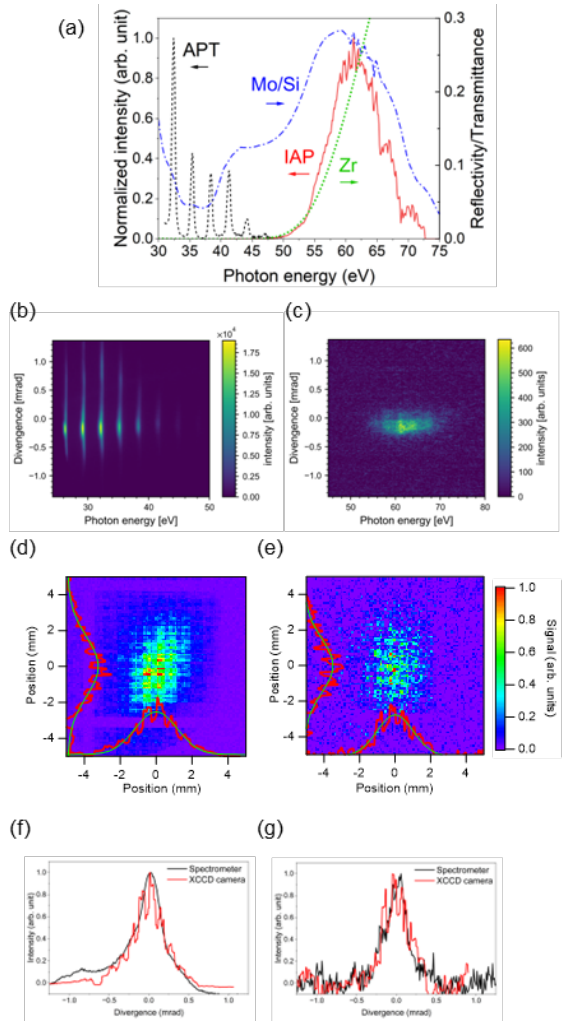

Fig. S1. Spectral and spatial characteristics of the XUV radiation used in the focusing experiments. (a) Spectral conditions relevant to the focusing experiments. The APT (black dashed line) and IAP (red solid line) spectra at the XUV focus are shown, along with the transmission of a 100-nm-thick Zr thin-film filter (green dotted line) and the reflectivity of the Mo/Si multilayer coating for s-polarized light (blue dashed–dotted line). (b) and (c) 2D spatial–spectral intensity distributions of XUV radiation for APT and IAP, respectively, measured downstream of the HHG source using an XUV spectrometer. (d) and (e) 2D beam profiles of XUV radiation for APT and IAP, respectively, measured after propagation through the focusing system using an X-ray CCD camera. The projected one-dimensional profiles along the horizontal and vertical directions are overlaid within each panel. The grid patterns and stripe features in the images originate from the support mesh used to hold the metal thin-film filters. (f) and (g) Beam divergences along the vertical direction for APT and IAP, respectively. The vertical direction corresponds to the slit direction of the spectrometer. The corresponding FWHM divergence values for the spectrometer and X-ray CCD camera are 0.44 and 0.48 mrad, respectively, for (f) and 0.35 and 0.39 mrad, respectively, for (g).

## *Supplementary Note S2. Design of ellipsoidal focusing mirror*

The mirror geometry was selected to provide a sufficient numerical aperture for tight focusing while maintaining unobstructed access to the focal region for beam characterization in the present beamline configuration. The focusing performance of the mirror was evaluated using wave-optical simulations based on the Fresnel–Kirchhoff diffraction integral. In these simulations, the XUV source was modeled with a spatial extent and divergence consistent with the experimental HHG conditions, with a divergence of approximately 0.23 mrad (FWHM of the angular distribution). The effect of surface figure errors on the focal spot size was investigated by incorporating model surface height variations representing different spatial wavelength bands into the phase of the reflected wavefront. Fig. S2 summarizes the simulation results, which show the calculated 2D focal intensity distributions for increasing levels of surface height variations. The simulations further reveal the influence of surface height errors in different spatial wavelength bands. As shown in Fig. S2, surface variations in the submillimeter spatial wavelength range (0.1–0.3, 0.3–1 mm) have a pronounced impact on the focal intensity distribution, resulting in significant focal spot broadening and distortion. In contrast, longer spatial wavelength components (1–3 mm) primarily modify the beam shape with a comparatively weaker effect on the focal confinement. The simulation results show that submicrometer focusing can be achieved when the surface figure error is sufficiently small, typically on the order of a few nanometers root mean square (RMS) or below for spatial wavelengths longer than 0.1 mm. In contrast, increased surface height variations resulted in the pronounced degradation of the focal spot, accompanied by the emergence of extended intensity distributions surrounding the central peak. These results provide a quantitative guideline for the allowable surface figure error required to achieve submicrometer focusing and serve as a practical design criterion for mirror fabrication in the present focusing scheme.

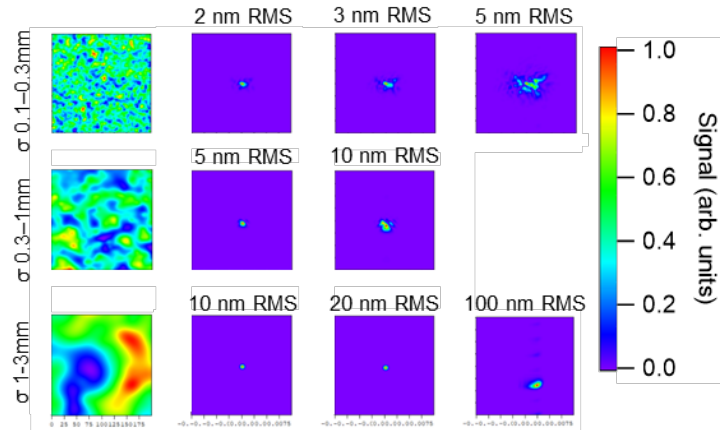

Fig. S2. Wave-optical simulations of the focal intensity distribution with different spatial wavelength components in the presence of surface height errors. The rows correspond to spatial wavelength ranges of 0.1–0.3 mm (top), 0.3–1 mm (middle), and 1–3 mm (bottom). In each row, the left panel shows a representative randomly generated surface height map of the ellipsoidal mirror containing the specified spatial wavelength band, and the right three panels show the corresponding focal intensity distributions calculated for progressively increasing RMS values. The horizontal axis of the surface map corresponds to the meridional coordinate ($\theta = 0.174$–$0.210$ rad), and the vertical axis corresponds to the sagittal direction over a 15° azimuthal range, corresponding to a physical surface extent of approximately 15–20 mm. The focal intensity distributions are plotted over a spatial window of 20 μm × 20 μm in the focal plane. The color scale represents the normalized intensity.

***Supplementary Note S3. Surface characterization of ellipsoidal mirror***
The surface quality of the fabricated ellipsoidal mirror was measured at multiple spatial scales (Fig. S3). Mirror surface metrology was performed after substrate fabrication and before multilayer coating deposition. The RMS values corresponding to different spatial wavelength ranges are 1.456 nm for spatial wavelengths larger than 3 mm, 0.731 nm for the 1–3-mm band, and 0.892 nm, evaluated over a 1 mm × 1 mm field of view, where submillimeter spatial wavelength components dominate. Considering the beam footprint on the mirror, these results indicate that the surface quality within the illuminated region is consistent with the requirements suggested based on the wave-optical simulations presented in Supplementary Note S2. The influence of these surface height variations on the experimentally obtained focal spot size is further discussed in Sec. 3.2.2.

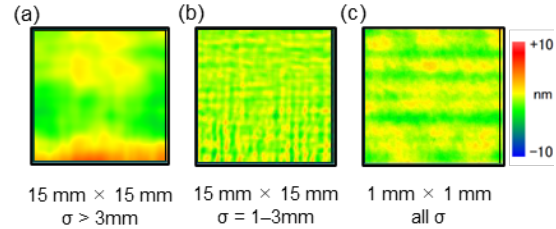

Fig. S3. Measured surface quality of fabricated ellipsoidal mirror at different spatial scales. (a) Surface height variations across the entire effective area (15 mm × 15 mm) for spatial wavelengths larger than 3 mm. (b) Band-pass-filtered surface height variations for spatial wavelengths of 1–3 mm extracted from the same area. (c) Surface height variations corresponding to submillimeter spatial wavelength components measured over a 1 mm × 1 mm field of view.

## *Supplementary Note S4. Beamline details*
Additional details of the beamline configuration are summarized below.
- The polarization of the driving laser pulse is rotated by 90° using a periscope, enabling attenuation of the infrared component at a grazing incidence of ~70° on the NbN-coated mirrors while maintaining high XUV throughput.
- The infrared beam reflected by the 500-nm-thick Al thin-film filter remains approximately collimated over the propagation distance to the PSD without significant spatial spreading.
- The 500-nm-thick Al filter exhibits negligible transmission in the infrared region under the present experimental conditions and a transmission of approximately 30%–50% in the 20–70-eV range.

## *Supplementary Note S5. Knife-edge acquisition parameters*
**Acquisition parameters for knife-edge measurements:**
For all measurements, the median of repeated exposures at each knife position was used as the robust estimator of the transmitted intensity. The acquisition parameters for each measurement condition are summarized in Table S1.

Table S1. Knife-edge acquisition parameters for APT and IAP measurements.

| Measurement condition | Pulses per exposure | Exposures per position | Scan step (μm) | Scan range (μm) |
|---|---|---|---|---|
| APT (Figs. 3(a) and (b)) | 1 | 20 | 0.1 | 3 |
| IAP (larger-spot, Figs. 3(c) and (d)) | 5 | 20 | 0.25 | 6 |
| IAP (sub-μm, Figs. 3(e) and (f)) | 10 | 10 | 0.15 | 2.4 |

These acquisition parameters were selected to ensure stable signal acquisition during the IAP operation, where the photon flux was reduced due to spectral selection. In addition, IAP generation based on multicolor waveform synthesis is sensitive to the spatial overlap of the driving fields, thereby increasing shot-to-shot fluctuations. Such fluctuations can also induce variations in the effective HHG source position, which may translate into fluctuations in the XUV focal position.

**Reproducibility of retrieved focal spot sizes:**
For the IAP measurements, three independent scans were performed in each transverse direction. The corresponding median-based FWHM focal spot sizes are summarized in Table S2.

Table S2. Reproducibility of focal spot sizes obtained from independent knife-edge scans, with horizontal and vertical values measured separately along each axis.

| Measurement condition | Direction | Scan 1 (μm) | Scan 2 (μm) | Scan 3 (μm) |
| --- | --- | --- | --- | --- |
| IAP (larger-spot, Figs. 3(c), (d)) | x | 1.54 ± 0.22 | 1.24 ± 0.28 | 1.51 ± 0.39 |
|  | y | 1.93 ± 0.36 | 1.43 ± 0.29 | 1.83 ± 0.34 |
| IAP (sub-μm, Figs. 3(e), (f)) | x | 0.46 ± 0.13 | 0.82 ± 0.25 | 0.97 ± 0.43 |
|  | y | 0.79 ± 0.21 | 0.36 ± 0.12 | 0.79 ± 0.24 |

*Supplementary Note S6. Effect of statistical estimators under beam-position fluctuations*
The influence of statistical estimators on knife-edge beam reconstruction under residual beam-position fluctuations was evaluated using numerical simulations. A Gaussian beam with a $1/e^2$ diameter of 1 μm (corresponding to a FWHM of 0.59 μm) was used, with random beam-position fluctuations characterized by a standard deviation of 1 μm. The beam size is expressed in terms of the $1/e^2$ diameter for a direct comparison with the amplitude of beam-position fluctuations.

At a representative knife position of −0.5 μm, the transmitted intensity was calculated for multiple realizations (50 samples) (Fig. 4(a)). Simulated transmitted intensities are plotted against the fluctuating beam position for a fixed knife position, along with the corresponding histograms of beam-position fluctuations (top) and transmitted intensities (right). Although the beam-position fluctuations are symmetric, the resulting distribution of transmitted intensities is asymmetric. In this example, the ideal transmission without fluctuations is 0.023, whereas the mean and median intensities are 0.354 and 0.061, respectively.

Repeating this procedure over all knife positions preserves the overall error function shape but increases its apparent width when the mean is used (Fig. 4(b)). The resulting beam reconstruction yields an apparent $1/e^2$ diameter of 3.97 ± 0.07 μm, which is significantly greater than the input value. In contrast, the median-based reconstruction recovers a diameter of 1.11 ± 0.05 μm. These results demonstrate that the median estimator provides a reliable and robust measure of the beam size under realistic experimental fluctuations.

*Supplementary Note S7. Mitigation of short-term focal drifts during scans*
Although the beam-position stabilization system suppresses long-period drift at the PSD reference point, the PSD monitors a proxy beam derived from the driving laser, which propagates along a path different from that of the XUV radiation. Consequently, mechanical perturbations may introduce relative shifts between the PSD-locked position and the actual XUV focal position. Such relative shifts can be observed experimentally even under active stabilization even though the PSD signal remains stable. Therefore, the effective XUV focus may shift slightly during the measurement. Because a single knife-edge scan requires sequential measurements at multiple knife positions over a timescale of a few tens of minutes, any shifts in the effective XUV focus during this period can change the apparent center of the knife-edge transition, resulting in a biased reconstruction of the beam size.

To ensure that the beam-size evaluation was performed under consistent conditions, we implemented a two-step scanning protocol. A coarse scan was initially performed to identify the approximate beam center over a range greater than the expected focal size, followed by a fine scan with submicrometer resolution around the focus. Only datasets for which the center position remained consistent within the resolution of the coarse scan were retained for the final beam-size reconstruction, and measurements showing larger deviations were excluded. This procedure effectively verifies focal position stability during each scan, ensuring that the extracted beam size reflects intrinsic focusing performance instead of temporal variations of the focal position.

### *Supplementary Note S8. Wave-optical simulation of local mirror surface variations*
The influence of local mirror surface variations on the focal spot size was investigated using wave-optical simulations based on experimentally measured surface data. The wavefield reflected from the mirror surface was propagated to the focal plane using the Rayleigh–Sommerfeld diffraction integral, numerically implemented with the chirp z-transform method. The surface height map measured over a 1 mm × 1 mm region of the ellipsoidal mirror was used as the input (Fig. S3(c)). Because it is not practical to measure the entire 15 mm × 15 mm mirror surface with submillimeter spatial resolution, the measured region was assumed to be representative of a typical local surface structure within the illuminated region.

To generate larger-area model surfaces suitable for wave-optical simulations, the measured surface map was resampled and extended to a 5 mm × 5 mm region while preserving its statistical properties, including the power spectral density. To model variations in illumination position, additional surface maps were generated by uniformly scaling the amplitude of the measured surface height variations. This represented different levels of local surface variations primarily arising from submillimeter spatial wavelength components within the measured region. For each surface map, the divergence of the incident XUV beam was set to 0.35 mrad (FWHM), based on the spectrometer measurement of the IAP beam shown in Fig. S1(c). The reflected wavefront was propagated to the focal plane to obtain the 2D intensity distribution. Subsequently, knife-edge signals were simulated by integrating the transmitted intensity as a function of knife position, and the corresponding beam sizes were reconstructed by fitting the resulting profiles using error functions.